\documentclass[fleqn,10pt]{wlscirep}
\usepackage[utf8]{inputenc}
\usepackage[T1]{fontenc}
\usepackage{svg}
\title{Ion solvation in atomic baths: from snowballs to polarons}

\author[1,2]{Saajid Chowdhury}
\author[1,2,*]{Jes\'us P\'erez-R\'ios}
\affil[1]{Department of Physics and Astronomy, Stony Brook University, Stony Brook, 11794, USA}
\affil[2]{Institute for Advanced Computational Science, Stony Brook University, Stony Brook, NY 11794, USA}

\affil[*]{jesus.perezrios@stonybrook.edu}

\begin{abstract}
Solvation, the result of the complicated interplay between solvent-solute and solvent-internal interactions, is one of the most important chemical processes. Consequently, a complete theoretical understanding of solvation seems a heroic task. However, it is possible to elucidate fundamental solvation mechanisms by looking into simpler systems, such as ion solvation in atomic baths. In this work, we study ion solvation by calculating the ground state properties of a single ion in a neutral bath from the high-density regime to the low-density regime, finding common ground for these two, in principle, disparate regimes. Our results indicate that a single $^{174}$Yb$^+$ ion in a bath of $^{7}$Li atoms forms a coordination complex at high densities with a coordination number of 8, with strong electrostriction, characteristic of the snowball effect. On the contrary, treating the atomic bath as a dilute quantum gas at low densities, we find that the ion-atom interaction's short-range plays a significant role in the physics of many-body bound states and polarons. Furthermore, in this regime, we explore the role of a putative ion trap, which drastically affects the binding mechanism of the ion and atoms from a quantum gas. Therefore, our results give a novel insight into the universality of ion-neutral systems in the ultracold regime and the possibilities of observing exotic many-body effects. 
\end{abstract}
\begin{document}

\flushbottom
\maketitle
%

\thispagestyle{empty}

\section*{Key Points}

\begin{itemize}

\item A global study of ion solvation in atomic baths from the high to the low-density regime.

\item The ion-atom short-range interaction is critical to understanding the presence of many-body bound states and polarons.  

\item The ion trapping potential drastically impacts many-body bound states and polaron formation.

\end{itemize}

\section*{Introduction}

The unique properties of charged-neutral interactions make a single ion in a neutral bath a testbed for cold chemistry~\cite{Perez-Rios2020,COTE20166,JPR2023,RevModPhysatomion,LOUS202265}, cluster physics\cite{Tomas2020}, solvation dynamics~\cite{Albrechtsen2023}, and many-body physics~\cite{bipolaron,astrakharchik2020ionic,Bruun,Michael,Meso1,Meso3}. When an ion is submerged in a sea of neutrals, the ion induces a dipole moment on the atoms that translates into the interaction $V(R)=-\frac{\alpha}{2R^4}$, where $\alpha$ is the polarizability of the atom and $R$ is the atom-ion distance. This interaction is much stronger than typical van der Waals interactions $\propto R^{-6}$. In this scenario, a single ion could bind to many atoms of the neutral bath, and depending on the nature of the bath, coordination complexes~\cite{CoordinationComplexes,CoordinationChemistryBook}, snowballs~\cite{snowball,snowball3,snowball2}, or polarons~\cite{Bruun,astrakharchik2020ionic} appear. 

In the case of a quantum gas, the bath can dress the atom-ion interaction, leading to a bound state between the ion and several atoms when the atom-ion interaction does not support any bound state, known as an ionic polaron, after the work of Landau and Pekar~\cite{Landau,Pekar1,Pekar2}. The ionic polaron shows two branches: one attractive, in which the impurity attracts atoms, and one repulsive, in which the impurity repels atoms despite showing a pure attractive two-body interaction~\cite{Bruun}. Nevertheless, despite the efforts to understand many-body effects in this novel paradigm, predictions on the properties of the many-body bound state and ionic polaron drastically differ throughout the literature without reaching a fulfilling explanation~\cite{astrakharchik2020ionic,Bruun,Michael,Schmelcher2017}. Similarly, in previous efforts, the role of the ion trap has never been included, even though most of the promising experimental platforms count on a tight ion trap. Along these lines, we have recently found that it is indeed possible to readily control the reactivity of the ion via the properties of the trap parameters~\cite{Henrik2023}. Therefore, the role of the trap on ion solvation in dilute quantum gases requires attention.

The very same charged-induced dipole interaction that gives rise to many-body effects is responsible for chemical reactions, such as three-body recombination~\cite{Mohammadi2021,Krukow2016,Perez-Rios2015,Perez-Rios2018,Review3BR,Weckesser2021}. In the case of an ion solvated in a helium droplet, the strong atom-ion interaction induces an electrostriction effect--the atoms are placed in their repulsive barrier, giving rise to snowballs~\cite{snowballAtkins}, as well as the formation of coordination complexes. On the contrary, when the solvent is very dilute, as is the case of quantum gases, it is generally assumed that the long-range interaction dominates the dynamics, and hence, short-range effects are negligible. However, this scenario is hardly compatible with ion solvation in helium droplets. 





In this paper, we unravel ion solvation in an atomic gas from the high-density regime (similar to helium droplets) to the low-density regime (dilute quantum gas). Our approach relies on quantum Monte Carlo techniques to calculate the ground state properties of the system. In particular, we study the impact of the short-range region of the ion-atom interaction potential on coordination complexes, snowballs, and many-body entities: many-body bound states and ionic polarons; thus, bridging the short-range (chemistry) with the long-range physics (many-body effects). Additionally, our study includes the trapping potential of the ion, exploring the role of the ion trap in the solvation.

\section*{Theoretical approach}

We consider a single $^{174}$Yb$^+$ ion in a sea of neutral $^7$Li atoms since it is one of the most promising candidates for reaching the quantum regime~\cite{Fedker2020}. Our simulations rely on Diffusion Monte Carlo (DMC), solving the N-body time-dependent Schr\"odinger equation (TDSE) in imaginary time, using the  Wick transformation, $\tau=\imath t$, yielding

\begin{equation}
\label{eq1}
\hbar \frac{\partial }{\partial \tau}|\psi(\mathbf{r}_1,\mathbf{r}_2,...,\mathbf{r}_n,\tau)\rangle=\sum_{i=1}^{n}\frac{\hbar^2 \nabla^2}{2m_i}|\psi(\mathbf{r}_1,\mathbf{r}_2,...,\mathbf{r}_n,\tau)\rangle- V(\mathbf{r}_1,\mathbf{r}_2,...,\mathbf{r}_N)|\psi(\mathbf{r}_1,\mathbf{r}_2,...,\mathbf{r}_n,\tau)\rangle.
\end{equation}
where $n$ the number of particles, $\hbar$ is the reduced Planck constant and $m_i$ stands for the mass of the i-th particle. Here, the potential energy landscape is built from pair-wise interactions as 

\begin{equation}
V(\mathbf{r}_1,\mathbf{r}_2,...,\mathbf{r}_n)=\sum_{i=2}^nV_{ai}(r_{i1})+\sum_{i=2}^n\sum_{j> i}^{n}V_{aa}(r_{ij})+V_{trap}(r_1),
\end{equation}
assuming that the particle with index $i=1$ is the ion. The trapping potential on the ion is assumed to be a harmonic trap given by $V_{trap}(r_1)=\frac{1}{2}m_1\omega^2 r_1^2$. The atom-atom interaction is taken as a Lennard-Jones potential, $V_{aa}(R)=\frac{C_{12}}{R^{12}}-\frac{C_{6}}{R^{6}}$, for the case of high atomic density, or as a hard sphere potential for the case of a neutral quantum bath, yielding the proper mean scattering length $\bar{a}$~\cite{Gribakin1993}. In this work, we employ two different atom-ion potentials. One is the generalization of the Lennard-Jones potential for charged-induced dipole interactions, reading as 
\begin{equation}
\label{pot1}
    V_{ai}(R)=\frac{C_8}{R^8}-\frac{C_4}{R^4},
\end{equation}
where $C_4=\alpha/2$ (in atomic units), $\alpha$ being the polarizability of the atom. The second atom-ion potential reads as
\begin{equation}
\label{pot2}
V_{ai}(R)=-C_4\frac{(R^2-c^2)}{(R^2+c^2)(b^2+R^2)^2},
\end{equation}
which is the most commonly-used for an ion  interacting with a dilute quantum gas~\cite{astrakharchik2020ionic,bipolaron,Bruun,Negretti}; it has the physical long-range interaction $\propto -C_4/R^4$ and a smooth behavior at short-range.

Next, noticing that Eq.~(\ref{eq1}) is a reaction-diffusion equation--a diffusion term associated with the kinetic energy and a sink or source term coming from the potential energy, we solve it by using a random walk with a branching process~\cite{DMC1,DMC2}. For a given number of particles $n$, the initial conditions for the single Yb$^+$ ion and $(n-1)$ Li atoms were determined by running one iteration of the basin-hopping algorithm~\cite{basin-hoping} to minimize the potential energy. All reported results converged with respect to the size of the time-step and the initial number of walkers. A detailed discussion on the method, convergence tests, and parameters used for the potentials can be found in the supporting information.

\section*{Results and Discussion}

\begin{figure}[t]
\centering
\includegraphics[width=0.7\linewidth]{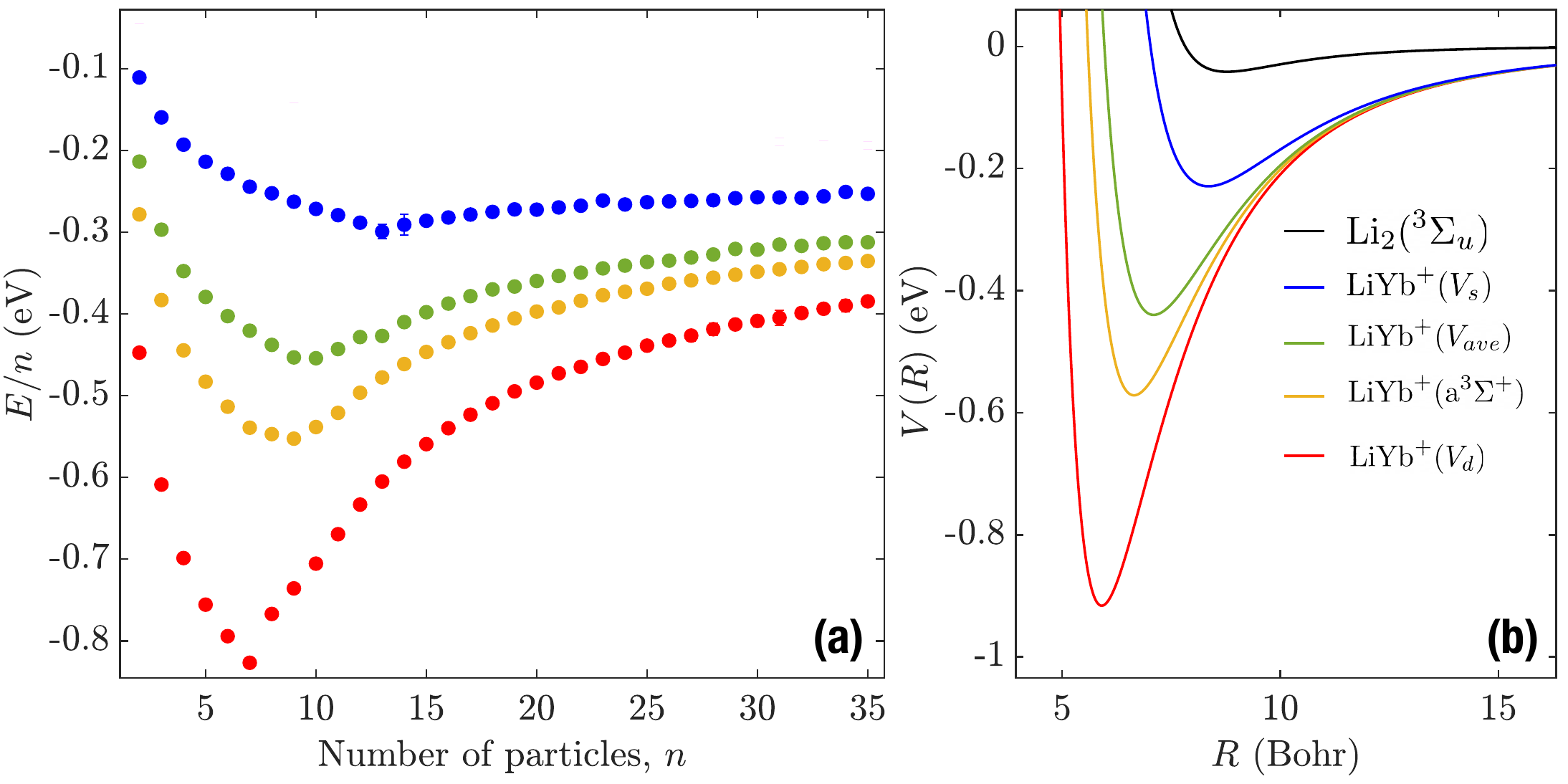}
\caption{ Ground state energy per particle of a single Yb$^+$ ion in a spin-polarized $^7$Li bath. Panel (a) displays the ground state energy per particle as a function of the number of particles. Panel (b) shows the atom-ion interaction potential employed in panel (a) following the same color scheme, along with the atom-atom potential. }
\label{fig1}
\end{figure}

An ion in solution binds to solvent molecules, forming solvation shells that \textit{cage} the ion from the rest of the solvent. In the case of an ion in an atomic bath, the solvation shell size and structure depend on the atom-ion and atom-atom interactions. Here, we investigate the ground state properties of a single $^{174}$Yb$^+$ ion in a sea of neutral $^7$Li atoms, and the results are displayed in panel (a) of Fig.~\ref{fig1}. It is worth emphasizing that the system under consideration is very different from ion solvation in helium droplets since the ion-atom interaction is, in our case, boosted by the large polarizability of Li--100-fold larger than He, and the interaction between the solvent, Li-Li, is 40-fold stronger than He-He. 

\begin{figure}[h!]
    \centering
    \includegraphics[width=0.8\linewidth]{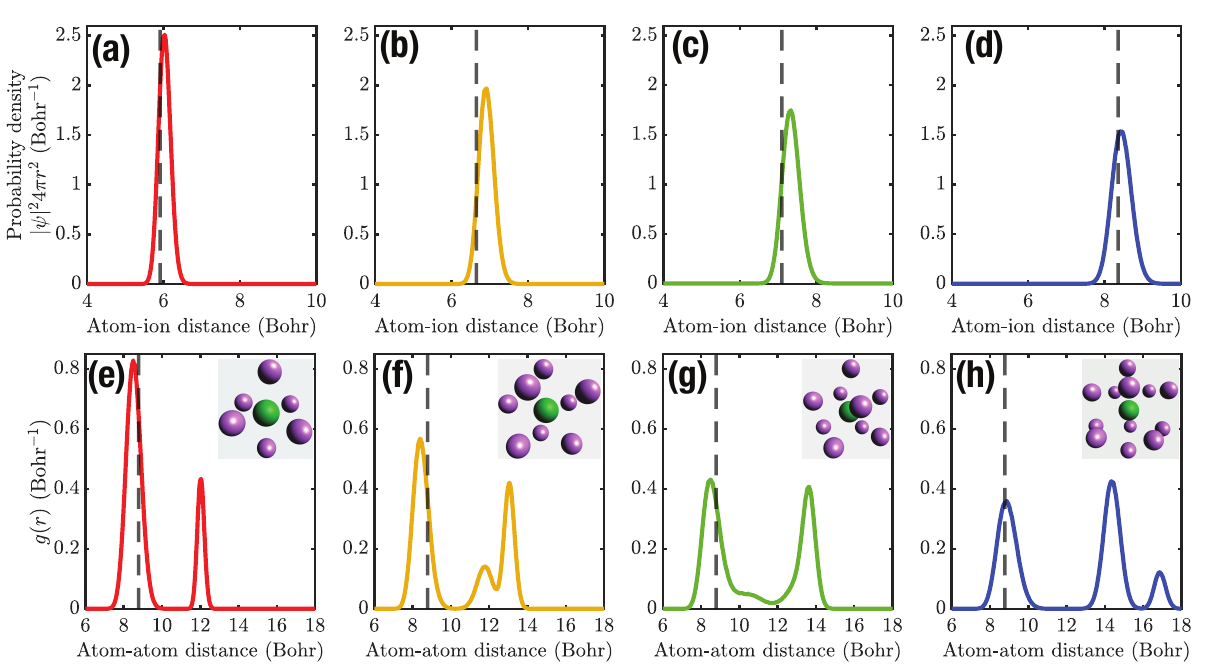}
    \caption{Radial atom-ion and atom-atom distributions of the first solvation shell. Panels (a) to (d) show the radial atom-ion distributions for the four potentials displayed in Fig.~\ref{fig1}. The dashed line stands for $R_e$ of the LiYb$^+$ potential. Panels (e) to (h) show the radial atom-atom distributions for the same potentials as in Fig.~\ref{fig1}. The dotted lines represent $R_e$ of the Li$_2$($^3\Sigma_u$) potential. The inset in each panel corresponds to the most probable atom and ion positions. The color code in this figure corresponds to the one employed in Fig.~\ref{fig1}.}
    \label{fig2}
\end{figure}

Fig.~\ref{fig1} presents the ground state energy (energy in short) per particle as a function of the number of particles for different atom-ion potentials, displayed in panel (b) of the same figure. The energy per particle first drops quickly, reaching a minimum; and from there, it grows up to acquiring an almost flat behavior for large numbers of particles. The minimum energy per particle represents the most stable cluster size, also known as the magic number, corresponding, in this case, to the first solvation shell and labeled as $N_S$. The Li-Li interaction potential is described by a Lennard-Jones potential satisfying the dissociation energy of the Li$_2$($^3\Sigma_u$) electronic state~\cite{Li2}, assuming that the atomic cloud is spin-polarized, for the sake of simplicity. We notice that the LiYb$^+$(a$^3\Sigma^+$) potential shows $N_S=9$ meaning that 8 atoms accrete around the ion, establishing the first solvation shell. However, the Li-Yb$^+$ interaction could also occur through the LiYb$^+$(A$^1\Pi$) potential. Hence, we also consider the average potential $V_{ave}=\frac{1}{4}($A$^1\Pi)+\frac{3}{4}($a$^3\Sigma^+$), showing a solvation shell of 10 particles (9 atoms), very close to the case of the pure triplet state interaction. We have not observed indications of a second solvation shell in both cases. The reason behind this is the strong atom-ion potential and the large zero-point energy of our system related to the light mass of Li.

To explore further the role of the atom-ion interaction potential on the solvation shell, we have designed a deeper potential, labeled by $V_d$, and a shallower potential, labeled by $V_s$, as shown in panel (b) of Fig.~\ref{fig1}. It is worth pointing out that all  LiYb$^+$ potentials show the same long-range coefficient $C_4$. The deeper potential shows a solvation shell of 7 particles (6 atoms) and the shallow one shows 13 particles (12 atoms) in the first solvation shell. These results seem to indicate that a shallow atom-ion potential leads to larger solvation shells. In other words, a shallow atom-ion potential can bind more atoms. To rationalize this observation, let us assume that the accretion process of the first solvation shell is isotropic, so the atoms spread themselves out evenly over a single spherical shell of radius $R_e$, the ion-atom equilibrium distance. The number of atoms in the solvation shell is proportional to the surface area, i.e.,  $N_S \propto {R_e}^2$, and for the potential at hand we find $R_e=\left(2C_8/C_4\right)^{1/4}$. Next, taking into account that $D_e=-V_{ai}(R_e)=C_4^2/4C_8$, where $D_e$ is the dissociation energy, we find $N_S \propto 1/\sqrt{D_e}$, explaining qualitatively our findings.

\begin{figure}[h]
\centering
\includegraphics[width=0.7\linewidth]{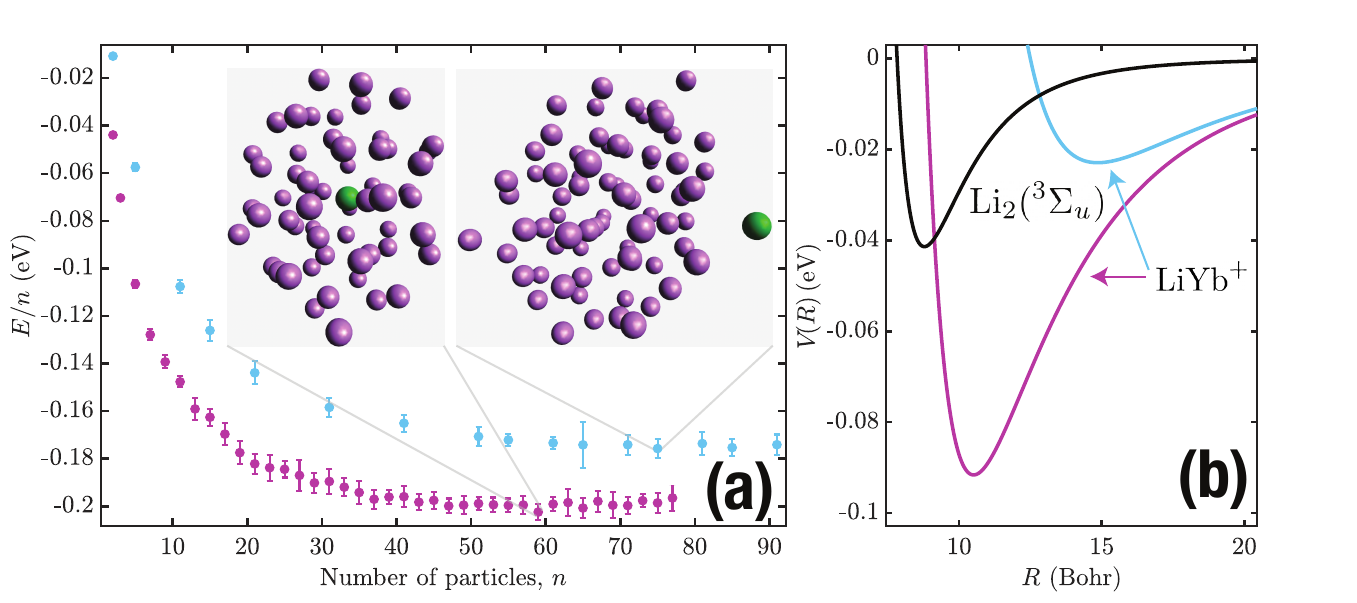}
\caption{Ground state energy per particle of a single Yb$^+$ ion in a $^7$Li bath. Extending Figure \ref{fig1}, panel (a) also shows the energy per particle as a function of the number of particles. Panel (b) shows the corresponding atom-ion potentials, as well as the Li$_2(^3\Sigma_u)$ in black.}
\label{fig3}
\end{figure}

Fig.~\ref{fig2} shows the radial atom-ion distribution and the atom-atom pair correlation function $g(R)$ associated with the first solvation shell of the LiYb$^+$ potentials presented in Fig.~\ref{fig1}. The radial atom-ion distribution is unimodal, meaning that the distances from the ion to all of the Li atoms are equivalent in the first solvation shell. However, we notice that the width of the distribution grows with $N_S$, meaning that the Li atoms for shallow potentials are more delocalized. On the contrary, the atom-atom pair correlation function shows a multimodal character, denoting the existence of different preferable distances between the Li atoms within the solvation shell. For $V_d$, the atom-atom pair correlation function shows a bimodal distribution consistent with the tetrahedral packing structure, one of the most common structures associated with the coordination number 6. For the LiYb$^+$(a$^3\Sigma^+$) potential, the pair correlation function shows three peaks associated with a square antiprism, as expected for the coordination number 8. The pair correlation function for $V_{ave}$ shows a bimodal distribution resembling a tricapped trigonal antiprism structure, characteristic of coordination number 9. Finally, the pair correlation function for $V_s$ shows an icosahedron, as expected for coordination number 12. Therefore, the structure of the solvation shell is susceptible to the strength of the atom-ion interaction. Based on these results, Yb$^+$ in a Li bath will form a coordination complex with a coordination number 8 or 9, which is the first solvation shell. 



On the other hand, we notice that for the first solvation shell, independently of the underlying atom-ion interactions, there is a non-negligible probability that the Li-Li distance is smaller than $R_e$ of the Li$_2$($^3\Sigma_u$) potential. This effect is known as electrostriction: the ion-atom interaction overcomes the atom-atom interaction, pushing the atoms against the repulsive wall of the two-body potential. Such an effect is characteristic of snowballs. The first peak of the radial distribution of atoms corresponds to a density $\approx 7\times 10^{21}$cm$^{-3}$, very close to the density of liquid lithium $\approx 4.4\times 10^{22}$cm$^{-3}$~\cite{davison1968compilation}, indicating a relatively stiff structure. However, the second and third peaks denote a more like-gas behavior: more delocalized Li atoms around the ion, within the snowball. This effect is more clear in panel (g) of Fig.\ref{fig2}, where there is a probability to find the atoms between the two peaks.


\begin{figure}
\centering
\includegraphics[width=0.5\linewidth]{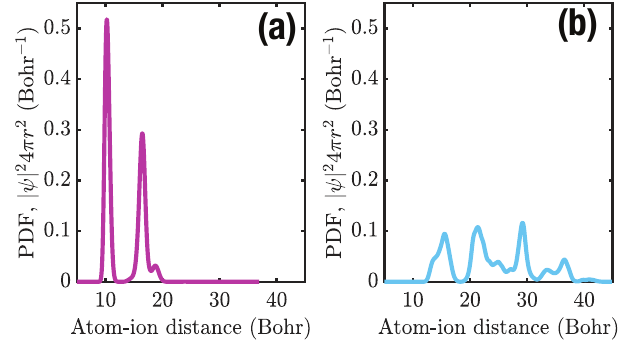}
\caption{Radial atom-ion distribution. }
\label{fig4}
\end{figure}

We have shown that $N_S$ depends on the depth of the atom-ion potential well: the shallower the potential, the larger the $N_S$. To explore this behavior further, we have calculated the energy per particle for two extremely shallow atom-ion potentials, whose details are given in the Supplemental information, and the results are shown in Fig.~\ref{fig3}. The energy per particle, after the expected quasi-linear dependence on the number of particles reaches a plateau, corresponds to a gas of atoms surrounding the ion. For atom-ion potentials deeper than the atom-atom interaction potentials, we observe a saturation of $N_S$ to approximately 60 atoms. However, when the atom-ion potential is shallower than the atom-atom interaction, $N_S=72$, but  the ion is not solvated anymore, and the atoms cluster away from the ion. In other words, if the ion-atom interaction is shallower than the atom-atom one, the ion is immiscible. This effect is better shown in Fig.~\ref{fig4}, where we report the radial atom-ion distributions. The figures reveal that when the atom-ion interaction is deeper than the atom-atom interaction, the ion shows two solvation shells. On the contrary, when the atom-ion potential is shallower than the atom-atom one, the radial atom-ion distribution shows a very spread-out and multi-modal distribution, which corresponds to an atomic cluster nearby the ion, corroborating the immiscibility of the ion when the atom-ion potential is shallower than the atom-atom one.

\subsection*{Low density regime: a dilute quantum bath}

Ion solvation in a low-density environment is commonly treated with a many-body approach. Consequently, the typical terminology employed in ion solvation at high densities is not commonly used. In addition, it is worth emphasizing that the typical energy scales of dilute quantum systems is $~\lesssim 1$mK$\equiv 8.617\times 10^{-8} $eV, and hence it is preferable to report energies in mK or $\mu$K. 

\begin{figure}[h!]
\centering
    \includegraphics[width=0.8\linewidth]{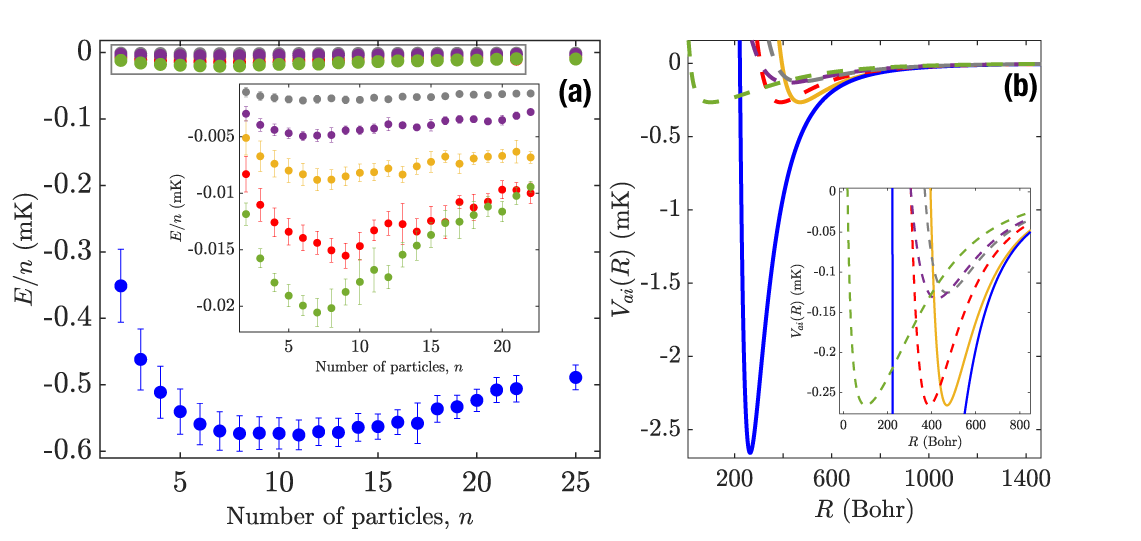}
\caption{Ground state energy of a single Yb$^+$ ion in a spin-polarized quantum gas of Li. Panel (a) displays the ground state energy per particle as a function of the number of particles. Panel (b) shows the different atom-ion potentials employed in panel (a). The potentials based on Eq.(\ref{pot1}) are represented by a solid line, whereas potentials based on Eq.(\ref{pot2}) are displayed with a dashed line. The inset shows a zoom-in on the region where most of the atom-ion potentials exhibit their minima. The color code in both panels is the same.}
\label{fig5}
\end{figure}

Two distinct scenarios have been predicted when an ion is placed in a dilute quantum gas. When the atom-ion potential supports bound states, the long-range atom-ion potential will bind many atoms to the ion, giving rise to many-body bound states~\cite{Meso1,Meso3}. In the language of ion solvation, the number of atoms in a many-body bound state is equivalent to the most stable cluster size in the system. For instance, in the case of Rb$^+$ ions in Rb atoms, the number of atoms bound to the ion is found to be $\sim 160$~\cite{astrakharchik2020ionic} or $\sim 40$~\cite{Bruun}. On the contrary, when the atom-ion potential does not support a bound state, the bath can dress the interaction, leading to the so-called polaron, a purely many-body effect that can only appear when the bath is a dilute quantum gas. Hence, it has no direct equivalent concept in standard ion solvation. Polarons have been predicted in the case of a Rb$^+$ ion in a bath of Rb atoms~\cite{Bruun,astrakharchik2020ionic}. However, the properties of ionic polarons in a mass-imbalanced system, in which the ion is much heavier than the atoms of the bath, remain unexplored even though it is the most common experimental scenario worldwide. In addition, in most experimental setups, the ion is generally trapped, and the effects of the trap in the formation of many-body bound states or polarons have yet to be addressed.

\begin{figure}[h!]
\centering
\includegraphics[width=0.4\linewidth]{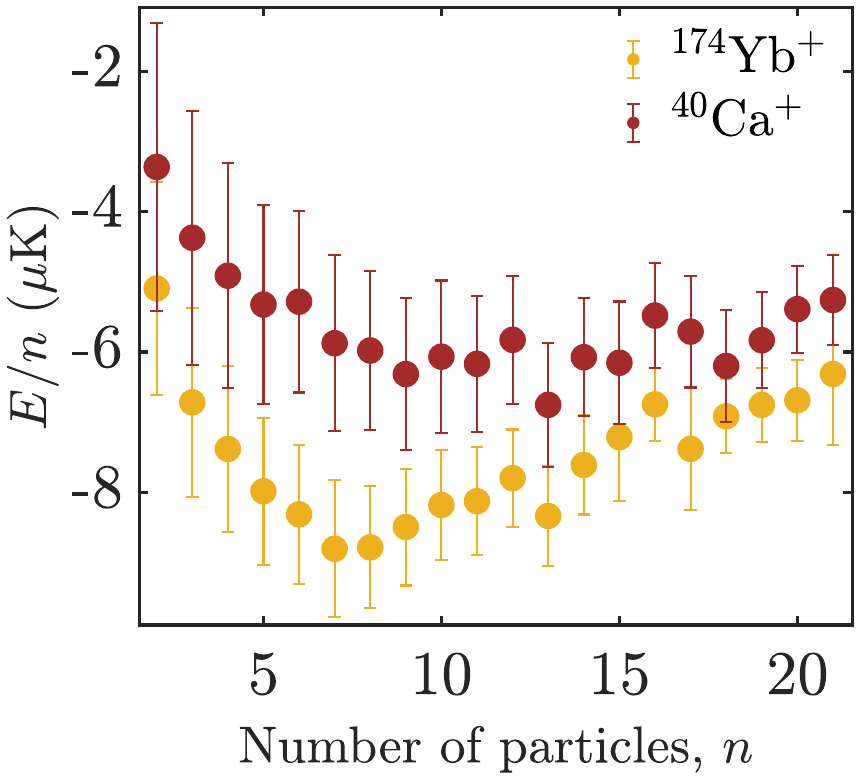}
\caption{The role of the ion mass on the formation of many-body bound states. In this case both of the atom-ion potentials show a single bound state.
}
\label{fig6}
\end{figure}


Here, we study the ground state properties of a single Yb$^+$ ion on a quantum bath of $^7$Li atoms, and the results for the energy per particle are shown in Fig.~\ref{fig5} for different atom-ion potentials. Each of these potentials supports a single bound state, showing the same long-range tail $-C_4/r^4$, but a different short-range region, as explained in the Supplemental information. Hence, since the binding mechanism is supposed to be due to the long-range tail of the potential~\cite{astrakharchik2020ionic}, all these potentials should show the same $N_S$. More importantly, the trend of the energy per particle as a function of the number of particles should be almost the same as well. However, our results indicate the contrary. The short-range region of the potential affects $N_S$, ranging between 5 and 8 atoms. In the same vein, we find that the short-range region of the potential affects the dependency of the energy per particle with the number of particles. Therefore, the short-range of the atom-ion potential plays a significant role in the number of atoms that the ion can bind within the many-body bound state regime, and this number is $\lesssim 10$.

To better understand the nature of our results, we have explored further the role of the mass of the ion in the ground state energy of the system. We assume that the atom-ion interaction is the same, but we change the mass of the ion, comparing Ca$^+$-Li to Yb$^+$-Li, and the results are displayed in Fig.~\ref{fig6}. The details of the potentials are elaborated in the Supplemental information. The energy per particle in the case of Yb$^+$-Li is systematically smaller than in the case of Ca$^+$-Li due to the reduced mass of the system. In addition, we notice that the ion's mass also affects the number of atoms the ion can bind, $N_S$ varies from 6 atoms to 8, for Yb$^+$-Li and Ca$^+$-Li, respectively. Therefore, as the ion mass approaches the atomic one, the size of the many-body bound state grows, explaining why, in the case of Rb$^+$-Rb the number of atoms in the many-body bound state is larger than the cases presented here.



\subsubsection*{Ion trap}
Ion-atom hybrid traps are the most common experimental platform for the study of fundamental ion-neutral processes in the low density regime. In ion-atom hybrid traps, the ions are trapped via a time-dependent potential. However, the full-quantum simulation of a trapped ion in a quantum bath is a heroic task. Nevertheless, ions can be trapped using a laser field, which gives rise to a time-independent harmonic trapping potential for the ion. In that case, a full quantum simulation is possible.

Here, we study the effects of the ion trap on the many-body bound states, and the results are shown in Fig.~\ref{fig7}. The interaction potentials employed in these calculations are shown in the Supplemental information. This figure shows the energy per particle as a function of the number of particles for different values of the trapping frequency on the ion. The results indicate that the trap has an appreciable effect on many-body bound states. As the trap frequency increases, overall, the energy per particle increases, and so does the position of the minimum of the ground state energy per particle. That is, the larger the trapping frequency, the larger the many-body bound state is. In the same vein, in the case of frequencies $\gtrsim $1~MHz, the many-body bound state survives even though the energy per particle is positive due to the harmonic trap. In other words, the many-body bound state becomes a trap state since the atoms are linked to the trapped ion. This observation aligns with our previous study on trap-induced atom-ion complexes~\cite{Henrik2023}, where we show that the trap properties affect the lifetime of atom-ion complexes, as well as chemical processes relying on those complexes, such as three-body recombination.

\begin{figure}[ht]
\centering
\includegraphics[width=0.5\linewidth]{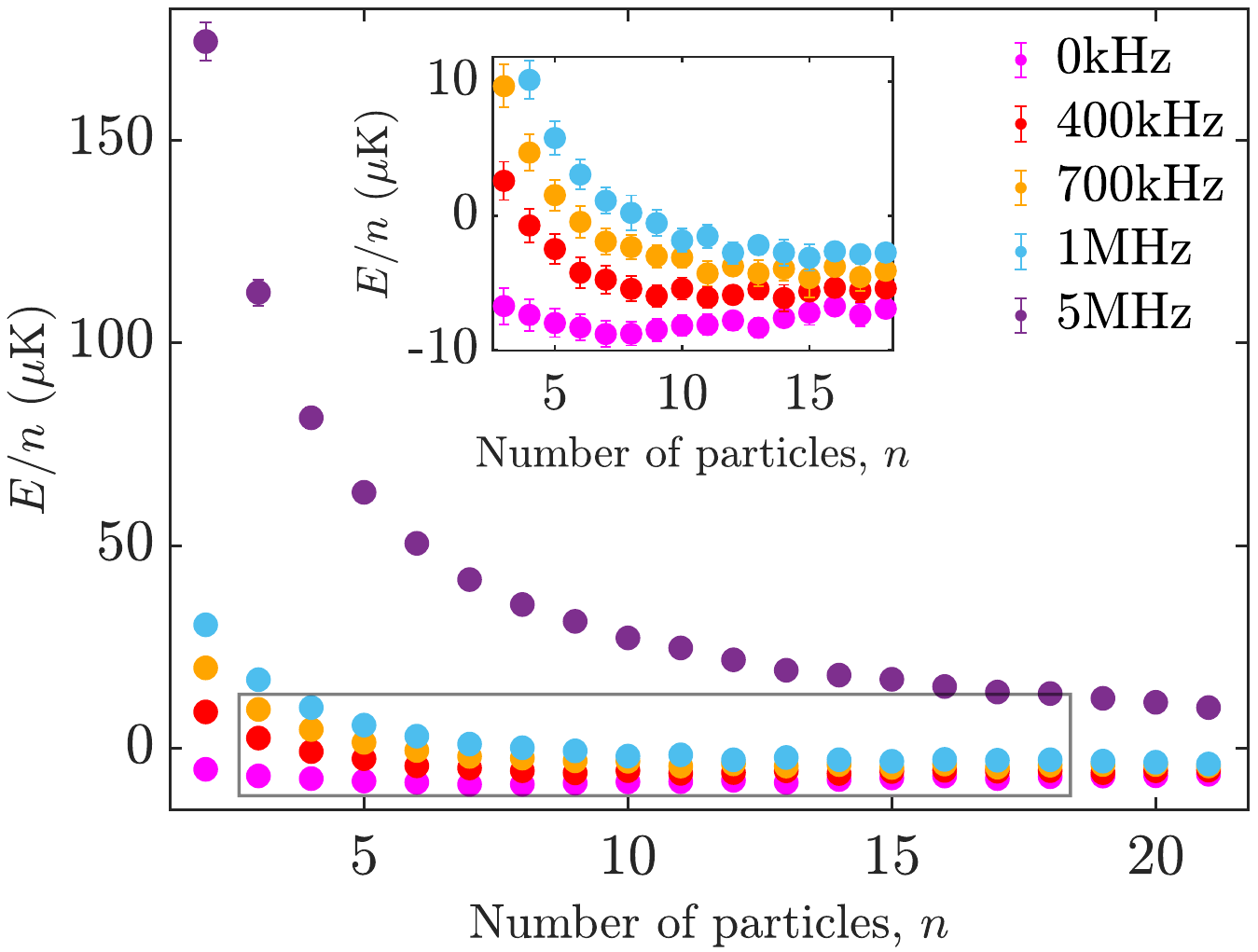}
\caption{The effect of the trap on the many-body bound state, in the low-density regime. The trap is modeled as a harmonic oscillator of a certain frequency centered at a fixed origin for the ion. The inset represents a zoom-in on the highlighted region.
}
\label{fig7}
\end{figure}

If the ion trap affects the properties of many-body bound states, it may affect the formation of polarons. So far, only trap-free studies on ionic polarons have been reported, and each of them reports a different number of atoms within the polaron~\cite{astrakharchik2020ionic,Bruun,Schmelcher2017}. Here, using our mass-imbalanced system, Yb$^+$-Li, we study the energy per particle as a function of the number of particles for different values of the trap frequencies, and the results are displayed in Fig.~\ref{fig8}. The details of the interaction potentials employed are shown in the Supplemental information. First, we notice a tremendous impact of the trapping frequency on the energy per particle regarding its trend, although showing a smooth effect on the polaron size. In detail, the larger the trap frequency, the less bound the ion-bath system, as we have seen in the case of many-body bound states (see Fig.~\ref{fig7}), independently of the number of particles; and the larger the polaron gets. The trapping potential on the ion helps the accretion toward the formation of polarons; however, the polaron energy diminishes. In other words, the trap dresses the atom-ion interaction, giving rise to a trapped weakly interacting polaron.



\begin{figure}[ht]
\centering
\includegraphics[width=0.4\linewidth]{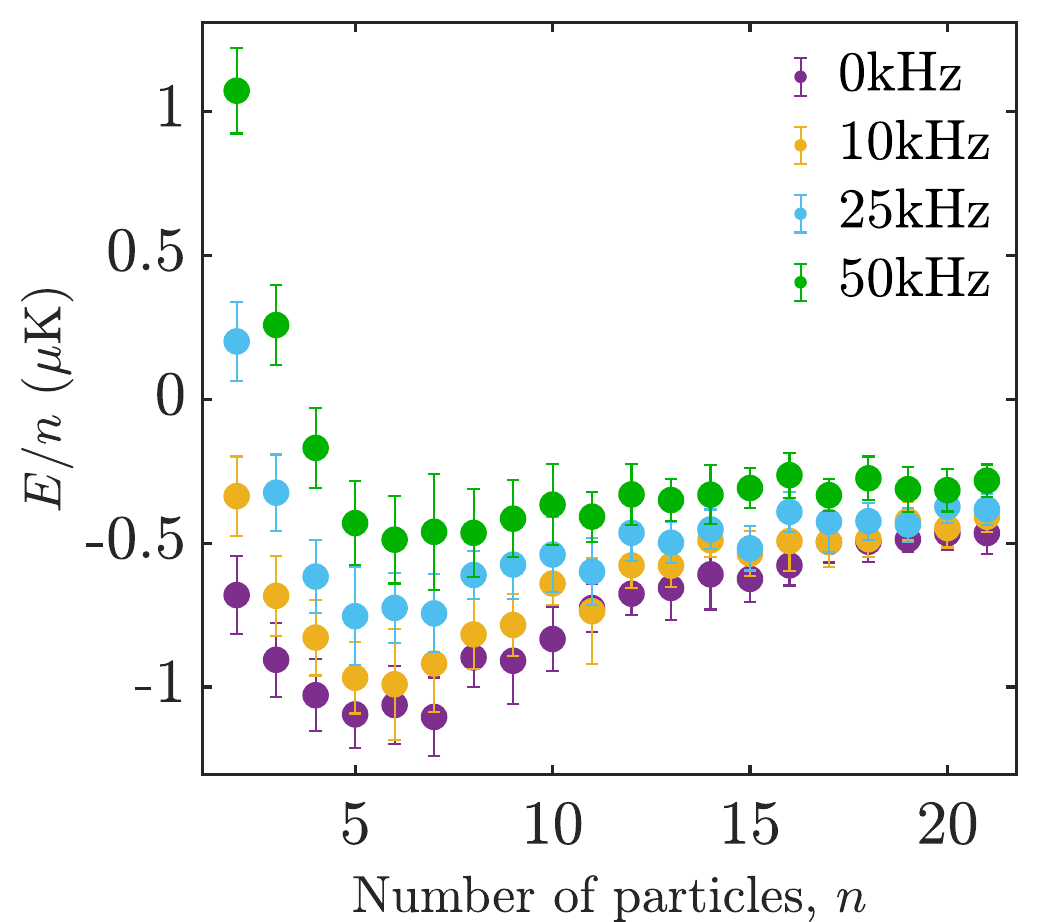}
\caption{Trap effect on the ionic polaron.}
\label{fig8}
\end{figure}

\section*{Conclusions}

Based on quantum Monte Carlo techniques, we presented a general study on the ground state properties of a single Yb$^+$ ion in Li solvent. In the case of a high-density solvent, we find that the first solvation shell is indeed a coordination complex of order 8 or 9. Moreover, the first solvation shell shows the direct effects of electrostriction, forming a snowball. The size of the solvation shell grows as the atom-ion potential gets shallower, reaching a point (when the atom-ion potential is shallower than the atom-atom one) in which the ion is immiscible in the solvent.  

In the case of a dilute quantum gas as a solvent, we have seen that the ion binds 5-10 atoms, forming many-body bound states, whose nature depends very much on the short-range atom-ion interaction. The many-body bound states we found are much smaller than predicted for Rb$^+$ in Rb. One possible explanation could be the much smaller reduced mass, or more considerable zero point energy, in our case, than in Rb$^+$ in Rb. Even when the ion-atom is unbound, we observe bound states associated with ionic polarons that generally trap 5-8 atoms, again, a smaller number than the ones predicted for Rb$^+$ in Rb. As with many-body bound states, our system's more considerable zero-point energy could explain these differences.
Furthermore, we have seen that many-body bound states and polarons are somewhat fragile clusters whose nature is easily molded by the ion trap. The trap frequency enhances the accretion process for many-body bound states and polarons, although showing a smaller binding energy. The trapping potential dresses the atom-ion interaction allowing for weakly bound states to survive on the trapping potential.

To summarize, ion solvation in an atomic solvent is an intriguing and exciting field of research in chemical physics. When the solvent density is high, it is well-known that the short-range interaction between the ion and the atoms affects the solvation dynamics. On the other hand, for low densities, it is believed that only the long-range region of the atom-ion interaction matters to describe the system's ground state. One possible solution to the dichotomy could be the existence of different solvation mechanisms depending on the density of the solvent. Another solution could be the fact that, indeed, short-range physics affects ion solvation in dilute gases, which, indeed, is the alternative that our results indicate. Therefore, our results elucidate the gap between ion solvation and impurity physics of an ion in a quantum bath.




\section*{Author contributions statements}
J. P.-R. envisioned the idea and supervised the project. S. C. conducted the calculations. Both authors contributed to the paper writing and figure making equally. J. P.-R. secured the funding. 




\section*{Acknowledgments}

The authors would like to thank the Simons Foundation for the support.



\section*{Conflict of interest}
The authors declare no conflict of interest.

\section*{Data availability statement}

All the data supporting our results are accessible at: \href{https://github.com/saajidchowdhury/Ionic-Polarons/blob/main/dmcPlots.m}{https://github.com/saajidchowdhury/Ionic-Polarons/blob/main/dmcPlots.m}. 

\section*{Ethical statement}
Not applicable.

\end{document}